\documentclass[11pt,draftcls,onecolumn]{IEEEtran}

\usepackage{enumerate}

\usepackage{amsbsy}
\usepackage{amssymb}
\usepackage{amscd}
\usepackage{amsmath}
\usepackage{epsf}
\usepackage{graphics}
\usepackage{graphicx}
\usepackage{epstopdf}
\usepackage{psfrag}
\usepackage{algorithm}
\usepackage{algorithmic}

\def\bx{{\mathbf{x}}}
\def\by{{\mathbf{y}}}
\def\bz{{\mathbf{z}}}
\def\bA{{\mathbf{A}}}
\def\bB{{\mathbf{B}}}
\def\bC{{\mathbf{C}}}
\def\bD{{\mathbf{D}}}

\def\bG{{\mathbf{G}}}
\def\bH{{\mathbf{H}}}
\def\bI{{\mathbf{I}}}

\def\bK{{\mathbf{K}}}

\def\bN{{\mathbf{N}}}
\def\bM{{\mathbf{M}}}
\def\bO{{\mathbf{O}}}

\def\bQ{{\mathbf{Q}}}

\def\bS{{\mathbf{S}}}

\def\bU{{\mathbf{U}}}

\def\bX{{\mathbf{X}}}

\def\b0{{\boldsymbol{0}}}

\newlength{\figwidth}
\newlength{\figwidthb}
\ifCLASSINFOpdf
\else
\fi

\begin{document}
\DeclareGraphicsExtensions{.eps}

\title{Solutions for the MIMO Gaussian Wiretap Channel with a Cooperative Jammer\thanks{The authors are with the Dept. of Electrical Engineering and Computer Science, University of California, Irvine, CA 92697-2625, USA. e-mail:\{afakoori, swindle\}@uci.edu}\thanks{This work was supported by the U.S. Army Research Office under the Multi-University Research Initiative (MURI) grant W911NF-07-1-0318.}}

\author{\normalsize S. Ali. A. Fakoorian*, {\it Student Member, IEEE} and A. Lee Swindlehurst, {\it Fellow, IEEE}}


\maketitle

\begin{abstract}
We study the Gaussian MIMO wiretap channel with a transmitter, a
legitimate receiver, an eavesdropper and an external helper, each
equipped with multiple antennas. The transmitter sends confidential
messages to its intended receiver, while the helper transmits jamming
signals independent of the source message to confuse the
eavesdropper. The jamming signal is assumed to be treated as noise at
both the intended receiver and the eavesdropper.  We obtain a
closed-form expression for the structure of the artificial noise
covariance matrix that guarantees no decrease in the secrecy capacity
of the wiretap channel.  We also describe how to find specific
realizations of this covariance matrix expression that provide good
secrecy rate performance, even when there is no non-trivial null space
between the helper and the intended receiver.  Unlike prior work, our
approach considers the general MIMO case, and is not restricted to
SISO or MISO scenarios.


\end{abstract}
\begin{center} \bfseries EDICS: WIN-CONT, WIN-PHYL, WIN-INFO, MSP-CAPC\end{center}

\newpage

\section{Introduction}
Recent information-theoretic research on secure communication has
focused on enhancing security at the physical layer. The wiretap
channel, first introduced and studied by Wyner \cite{Wyner}, is the
most basic physical layer model that captures the problem of
communication security. This work led to the development of the notion
of perfect secrecy capacity, which quantifies the maximum rate at
which a transmitter can reliably send a secret message to its intended
recipient, without it being decoded by an eavesdropper.  The Gaussian
wiretap channel, in which the outputs of the legitimate receiver and
the eavesdropper are corrupted by additive white Gaussian noise, was
studied in \cite{GW1978}. The secrecy capacity of a Gaussian wiretap
channel, which is in general a difficult non-convex optimization
problem, has been addressed and solved for in
\cite{Hassibi}-\cite{Bustin}.  The secrecy capacity under an average
power constraint is treated in \cite{KhistiMIMO} and \cite{ours},
where in \cite{KhistiMIMO} a beamforming approach, based on the
generalized singular value decomposition (GSVD), is proposed that
achieves the secrecy capacity in the high SNR regime. In \cite{ours},
we propose an optimal power allocation that achieves the secrecy
capacity of the GSVD-based multiple-input, multiple-output (MIMO)
Gaussian wiretap channel for any SNR. In \cite{Bustin}, a closed-form
expression for the secrecy capacity is derived under a certain
power-covariance constraint.

It was shown in \cite{Csiszar} that, for a wiretap channel without
feedback, a non-zero secrecy capacity can only be obtained if the
eavesdropper's channel is of lower quality than that of the intended
recipient. Otherwise, it is infeasible to establish a secure link
under Wyner's wiretap channel model. In such situations, one approach
is to exploit user cooperation in facilitating the transmission of
confidential messages from the source to the destination.  In
\cite{Tekin}-\cite{IFCHlp}, for example, a four-terminal
relay-eavesdropper channel is considered, where a source wishes to
send messages to a destination while leveraging the help of a
relay/helper node to hide the messages from the eavesdropper. While
the relay can assist in the transmission of confidential messages, its
computational cost may be prohibitive and there are difficulties
associated with the coding and decoding schemes at both the relay and
the intended receiver.  Alternatively, a cooperating node can be used as a
helper that simply transmits jamming signals, independent of the
source message, to confuse the eavesdropper and increase the range of
channel conditions under which secure communications can take
place. The strategy of using a helper to improve the secrecy of the
source-destination communication is generally known as cooperative
jamming \cite{Tekin,Petropulu} or noise-forwarding \cite{Gamal} in
prior work.

In \cite{Tekin}, the scenario where multiple single-antenna users
communicate with a common receiver (i.e., the multiple access channel) in the
presence of an eavesdropper is considered, and the optimal transmit
power allocation that achieves the maximum secrecy sum-rate ia
obtained. The work of \cite{Tekin} shows that any user prevented from
transmitting based on the obtained power allocation can help increase
the secrecy rate for other users by transmitting artificial noise to
the eavesdropper (cooperative jamming).  In \cite{Petropulu}, a
source-destination system in the presence of multiple helpers and
multiple eavesdroppers is considered, where the helpers can transmit
weighted jamming signals to degrade the eavesdropper's ability to
decode the source. While the objective is to select the weights so as to
maximize the secrecy rate under a total power constraint, or to
minimize the total power under a secrecy rate constraint, the results
in \cite{Petropulu} yield sub-optimal weights for both single
and multiple eavesdroppers, due to the assumption that
the jamming signal must be nulled at the destination. The noise
forwarding scheme of \cite{Gamal} requires that the interferer's
codewords be decoded by the intended receiver. A
generalization of \cite{Tekin,Petropulu} and \cite{Gamal} is proposed
in \cite{IFCHlp}, in which the helper's codewords do not have to be
decoded by the receiver.

The prior work in \cite{Tekin}-\cite{IFCHlp} assumes single antenna
nodes and models single-input, single-output (SISO) or multiple-input,
single-output (MISO) cases.  A more general MIMO case with multiple
cooperative jammers was studied in \cite{WangS}, in which the jammers
aligned their interference to lie within a pre-specified ``jamming
subspace'' at the receiver, but the dimensions of the subspace and the
power allocation were not optimized.  In this paper, we also address
the general MIMO case, where the transmitter, legitimate receiver,
eavesdropper and helper are in general all equipped with multiple
antennas. The transmitter sends confidential messages to its intended
receiver, while the helper node assists the transmitter by sending
jamming signals independent of the source message to confuse the
eavesdropper. While the previous work on this problem shows the
fundamental role of jamming as a means to increase secrecy rates, it
also emphasizes the fact that that non-carefully designed jamming
strategies can preclude secure communication \cite{Laneman}.

In this
work, we derive a closed-form expression for the structure of the
artificial noise covariance matrix of a cooperating jammer that
guarantees no decrease in the secrecy capacity of the wiretap channel,
assuming the jamming signal from the helper is treated as noise at
both the intended receiver and the eavesdropper.  We describe
algorithms for finding specific realizations of this covariance
expression that provide good secrecy rate performance, and show that
even when there is no non-trivial nullspace between the helper and the
intended receiver, the helper can still transmit artificial noise that
does not impact the mutual information between the transmitter and the
intended receiver, while decreasing the mutual information between the
transmitter and the eavesdropper. Hence, the secrecy level of the
confidential message is increased.  The situation we consider is
different from the one in \cite{GoelN08}, where the transmitter itself
rather than an external helper broadcasts artificial noise to degrade
the eavesdropper's channel.  However, both approaches are able to
achieve a positive perfect secrecy rate in scenarios where the secrecy
capacity in the absence of jamming is zero.

The remainder of the paper is organized as follows. In Section \ref{sysmod}, we
describe the system model for the helper-assisted Gaussian MIMO
wiretap channel and formulate the problem to be solved. In Sections \ref{analysis} and \ref{scenarios}, we derive the artificial noise covariance matrix that guarantees
no decrease in the secrecy capacity of the wiretap channel. Numerical
results in Section \ref{numrst} are presented to illustrate the proposed
solution. Finally, Section \ref{conc} concludes the paper.

\textbf{Notation:} Throughout the paper, we use boldface uppercase
letters to denote matrices. Vector-valued random variables are written
with non-boldface uppercase letters ({\em e.g.,} $X$), while the
corresponding lowercase boldface letter ($\bx$) denotes a specific
realization of the random variable. Scalar variables are written with
non-boldface (lowercase or uppercase) letters.  We use $(.)^T$ to
represent matrix transposition, $(.)^H$ the Hermitian (i.e.,
conjugate) transpose, Tr(.) the matrix trace, $E$ the expectation
operator, \textbf{I} the identity matrix, and \textbf{0} a matrix or
vector with all zeros.  Mutual information between the random variables $A$ and $B$ is
denoted by $I(A;B)$, and $\mathcal{CN}(0,1)$ represents the complex
circularly symmetric Gaussian distribution with zero mean and unit
variance.

\section{System Model}  \label{sysmod}

We consider a MIMO wiretap channel that includes a transmitter, an
intended receiver, a helping interferer and an eavesdropper, with
$n_t$, $n_r$, $n_h$ and $n_e$ antennas, respectively.  The transmitter
sends a confidential message to the intended receiver with the aid of
the helper, in the presence of an eavesdropper. We assume that the
helper does not know the confidential message and transmits only a
Gaussian jamming signal which is not known at the intended receiver
nor the eavesdropper and which is treated as noise at both receivers.
The mathematical model for this scenario is given by:
\begin{eqnarray}
\by_1 & = & \bH_1\bx_1+ \bG_2\bx_2 + \bz_1 \label{hlp1} \\
\by_2 & = & \bH_2\bx_2+ \bG_1\bx_1 + \bz_2 \; , \label{hlp2}
\end{eqnarray}
where $\bx_1$ is a zero-mean $n_{t}\times1$ transmitted signal vector,
$\bx_2$ is a zero-mean $n_{h}\times1$ jamming vector transmitted by
the helper, and $\bz_1\in\mathbb{C}^{n_{r}\times1}$,
$\bz_2\in\mathbb{C}^{n_{e}\times1}$ are additive white Gaussian noise
(AWGN) vectors at the intended receiver and the eavesdropper,
respectively, with i.i.d. entries distributed as $\mathcal{CN}(0,1)$.
The matrices $\bH_1,\bG_1$ represent the channels from the transmitter
to the intended receiver and eavesdropper, respectively, while
$\bH_2,\bG_2$ are the channels from the helper to the eavesdropper and
intended receiver, respectively.  The channels are assumed to be
independent of each other and full rank with arbitrary dimensions.  We
also assume that the transmitter has full channel state information
and is aware of the effective noise covariance at both receivers,
where the effective noise is the background noise plus the received
artificial noise. Both the helper and the eavesdropper are also aware
of all channel matrices as well.

The jamming signal transmitted by the helper satisfies an average power constraint:
\begin{equation}\label{hlp3}
\text{Tr}(E\{X_2X_2^H\})=\text{Tr}(\bK_w)\leq P_h
\end{equation}
where $X_2$ is the random variable associated with the specific
realization $\bx_2$ and $\bK_w$ is the corresponding covariance
matrix.  The channel input is subject to a matrix power constraint
\cite{Bustin,RLiu}
\begin{equation}\label{hlp4}
E\{X_1X_1^H\}=\bK_x\preceq \bS
\end{equation}
where $\bK_x$ is the input covariance matrix, $\bS$ is a positive
semi-definite matrix, and ``$\preceq$'' denotes that $\bS-\bK_x$ is
positive semi-definite.  Note that~(\ref{hlp4}) is a rather general
power constraint that subsumes many other important power constraints,
including the average total and per-antenna power constraints as
special cases.  The approach developed in this paper will assume that
$P_h$ and $\bS$ (or $\text{Tr}(\bS)\leq P_t$) are fixed, and that
power is not allocated jointly between the transmitter and helper.
The numerical results presented later, however, will illustrate the
trade-off associated with the power allocation when $P_h+P_t$ is fixed.

As mentioned before, we assume Gauusian signaling for the helper. Thus
the effective noise at both receivers is Gaussian and consequently
the above MIMO wiretap channel model is Gaussian. For this case, a
Gaussian input signal is the optimal choice \cite{TLiu,RLiu}. Hence,
the general optimization problem is equivalent to finding the matrices
$\bK_x\succeq0$ and $\bK_w\succeq0$ that allow the
secrecy capacity of the network to be obtained.  A matrix
characterization of this optimization problem is given by:
\begin{eqnarray}\label{hlp5}
C_{sec}&=&\max_{\bK_x\succeq0, \bK_w\succeq0} [I(X_1; Y_1)- I(X_1; Y_2)]\nonumber\\
&=&\max_{\bK_x\succeq0, \bK_w\succeq0} \log|\bK_x\bH_1^H(\bG_2\bK_w\bG_2^H+\textbf{I})^{-1}\bH_1+\textbf{I}|\nonumber\\
&& \qquad \qquad \quad -\log|\bK_x\bG_1^H(\bH_2\bK_w\bH_2^H+\textbf{I})^{-1}\bG_1+\textbf{I}| \; ,
\end{eqnarray}
where the non-convex maximization problem in carried out under the power
constraints given in~(\ref{hlp3}) and~(\ref{hlp4}).  $\\$
\textbf{Lemma 1:} For a given
$\bK_w$, the maximum of (\ref{hlp5}) is given by
\begin{equation}\label{hlp6}
C_{sec}(\bS)=\sum_{i=1}^{\rho}\log\gamma_i
\end{equation}
where $\gamma_i$, $i=1,\cdots,\rho$, are the generalized eigenvalues of the pencil
\begin{equation}\label{hlp7}
(\bS^{\frac{1}{2}}\bH_1^H(\bG_2\bK_w\bG_2^H+\textbf{I})^{-1}\bH_1\bS^{\frac{1}{2}}+\textbf{I},\quad \bS^{\frac{1}{2}}\bG_1^H(\bH_2\bK_w\bH_2^H+\textbf{I})^{-1}\bG_1\bS^{\frac{1}{2}}+\textbf{I})
\end{equation}
that are greater than 1.$\\$
\textbf{Proof:} When the optimization problem in (\ref{hlp5}) is performed over
$\bK_x$ under the matrix power constraint (\ref{hlp4}) for a given $\bK_w$, it
is equivalent to a simple MIMO Gaussian wiretap channel without
a helper, where the noise covariance matrices at the receiver and the
eavesdropper are $(\bG_2\bK_w\bG_2^H+\textbf{I})$ and
$(\bH_2\bK_w\bH_2^H+\textbf{I})$, respectively. The above lemma is a
natural extension of \cite{Bustin} and \cite[Theorem 3]{RLiu} for the
standard MIMO Gaussian wiretap channel.

Note that since both elements of the pencil~(\ref{hlp7}) are strictly
positive definite, all of the generalized eigenvalues are real and
positive \cite{RLiu,Horn}.  In~(\ref{hlp6}), a total of $\rho$ of them
are assumed to be greater than one. Clearly, if there are no such
eigenvalues, then the information signal received at the intended
receiver is a degraded version of that of the eavesdropper, and in this
case the secrecy capacity is zero.  Note also that Lemma~1 only
provides the secrecy capacity for the optimal $\bK_x$, but does not
give an explicit expression for this $\bK_x$.  A general expression
for the maximizing $\bK_x$ will be given in the next section.

To solve the general optimization problem in (\ref{hlp5}), we would
need to find the $\bK_w$ that maximizes~(\ref{hlp6}). Unfortunately,
this appears to be a very difficult problem to solve without resorting
to some type of {\em ad hoc} search.  In the following we obtain a
sub-optimal closed-form solution for the artificial noise covariance
matrix $\bK_w$ that guarantees no decrease in the mutual information
between the transmitter and the intended receiver compared with the case where
$\bK_w=\mathbf{0}$, while maintaining the power constraint
in~(\ref{hlp5}).  Hence, the new non-zero $\bK_w$ will only interfere
with the eavesdropper, and the secrecy level of the confidential
message will be increased. Once such a $\bK_w$ is found, additional
improvement in the secrecy rate can be achieved if the transmitter
updates its covariance matrix $\bK_x$ for the obtained $\bK_w$. The
final secrecy rate for this method is obtained by simply
computing~(\ref{hlp6}) and~(\ref{hlp7}) for the resulting
$\bK_w$. Note that we will not propose an iterative algorithm that
would further alternate between calculating $\bK_x$ and $\bK_w$. We
will see in the next section that there is no clear way to update
$\bK_w$ from a known non-zero value.

\section{Analytical Method}  \label{analysis}

We begin with the case where the helper transmits no signal
$(\bK_w=0)$. In this case, the communication system is reduced to a simple
MIMO Gaussian wiretap channel without helper.  Based on Lemma~1,
the maximum of (\ref{hlp5}) when $\bK_w=0$ is obtained by applying the generalized
eigenvalue decomposition to the following two Hermitian positive
definite matrices \cite{Bustin, RLiu}:
$$\bS^{\frac{1}{2}}\bH_1^H\bH_1\bS^{\frac{1}{2}}+\textbf{I},\quad \bS^{\frac{1}{2}}\bG_1^H\bG_1\bS^{\frac{1}{2}}+\textbf{I} \; .$$
In particular, there exists an invertible generalized eigenvector matrix $\bC$ such that \cite{Horn}
\begin{equation}\label{hlp8}
\bC^H\left[\bS^{\frac{1}{2}}\bG_1^H\bG_1\bS^{\frac{1}{2}}+\textbf{I}\right]\bC=\textbf{I}
\end{equation}
\begin{equation}\label{hlp9}
\bC^H\left[\bS^{\frac{1}{2}}\bH_1^H\bH_1\bS^{\frac{1}{2}}+\textbf{I}\right]\bC=\mathbf{\Lambda}
\end{equation}
where $\mathbf{\Lambda}=\text{diag}\{\lambda_1,...,\lambda_{n_t}\}$ is
a positive definite diagonal matrix and $\lambda_1,...,\lambda_{n_t}$
represent the generalized eigenvalues. Without loss of generality, we
assume the generalized eigenvalues are ordered as
$$\lambda_1\geq...\geq\lambda_b >1 \geq \lambda_{b+1}\geq...\geq\lambda_{n_t}>0$$
so that a total of $b$ $(0\leq b \leq n_t)$ are assumed to be greater than 1.
Hence, we can write $\mathbf{\Lambda}$ as
\begin{equation}\label{hlp10}
\mathbf{\Lambda}=\left[
\begin{array}{ccc}
\mathbf{\Lambda}_1 & 0\\
0 & \mathbf{\Lambda}_2
\end{array}
\right]
\end{equation}
where $\mathbf{\Lambda}_1=\text{diag}\{\lambda_1,...,\lambda_b\}$ and
$\mathbf{\Lambda}_2=\text{diag}\{\lambda_{b+1},...,\lambda_{n_t}\}$. Also,
we can write $\bC$ as
\begin{equation}\label{hlp11}
\bC=[\bC_1\quad \bC_2]
\end{equation}
where $\bC_1$ is the $n_t \times b$ submatrix representing the
generalized eigenvectors corresponding to
$\{\lambda_1,...,\lambda_b\}$ and $\bC_2$ is the $n_t \times (n_t-b)$
submatrix representing the generalized eigenvectors corresponding to
$\{\lambda_{b+1},...,\lambda_{n_t}\}$.

For the case of $\bK_w=0$, the secrecy capacity of (\ref{hlp5}) under the matrix power constraint (\ref{hlp4})
is given by (Lemma 1 or \cite[Theorem 3]{RLiu}):
\begin{equation}\label{cs}
C_{sec}=\sum_{i=1}^{b}\log\lambda_i=\log|\mathbf{\Lambda}_1|
\end{equation}
and the input covariance matrix $\bK_x^*$ that maximizes (\ref{hlp5}) is given by (\cite{Bustin,RLiu}):
\begin{equation}\label{hlp12}
\bK_x^*= \bS^{\frac{1}{2}}\bC \left[
\begin{array}{ccc}
(\bC_1^H\bC_1)^{-1} & 0\\
0 & 0
\end{array}
\right]\bC^H \bS^{\frac{1}{2}} \; .
\end{equation}
Note that~(\ref{hlp12}) is a general expression for the $\bK_x$ that
optimizes~(\ref{hlp5}) for a given $\bK_w$ even when $\bK_w\ne 0$, although in this case the
$\bC$ will be the generalized eigenvector matrix of the pencil~(\ref{hlp7}).
From~(\ref{hlp9}) we note that $\bH_1^H\bH_1$ can be written as
\begin{equation}\label{hlp13}
\bH_1^H\bH_1= \bS^{-1/2}\left[\bC^{-H} \left[
\begin{array}{ccc}
\mathbf{\Lambda}_1 & 0\\
0 & \mathbf{\Lambda}_2
\end{array}
\right]\bC^{-1}-\textbf{I}\right] \bS^{-1/2} \; . 
\end{equation}
The following lemma gives the mutual information $I(X_1;Y_1)$ between
the transmitter and the intended receiver when $\bK_w=0$ and $\bK_x$
is given by (\ref{hlp12}).$\\$
\textbf{Lemma 2:} The following equality holds:
\begin{equation}\label{hlp14}
I(X_1;Y_1)|_{\bK_w=0,\bK_x=\bK_x^*}=\log\left|\bK_x^*\bH_1^H\bH_1+\textbf{I}\right|=\log\left|(\bC_1^H\bC_1)^{-1}\mathbf{\Lambda}_1\right| \; .
\end{equation}
\textbf{Proof:} Following the same steps as the proof of
\cite[App. D]{Bustin} and using (\ref{hlp12}) and (\ref{hlp13}), we have
\begin{eqnarray}
\left|\bK_x^*\bH_1^H\bH_1+\textbf{I}\right|&=&\left|\bS^{\frac{1}{2}}\bC \left[
\begin{array}{ccc}
(\bC_1^H\bC_1)^{-1} & 0\\
0 & 0
\end{array}
\right]\bC^H\times\left[\bC^{-H} \left[
\begin{array}{ccc}
\mathbf{\Lambda}_1 & 0\\
0 & \mathbf{\Lambda}_2
\end{array}
\right]\bC^{-1}-\textbf{I}\right] \bS^{-1/2}+\textbf{I}\right|\nonumber \\
&=&\left|\left[
\begin{array}{ccc}
(\bC_1^H\bC_1)^{-1} & 0\\
0 & 0
\end{array}
\right]\times\left[
\begin{array}{ccc}
\mathbf{\Lambda}_1 & 0\\
0 & \mathbf{\Lambda}_2
\end{array}
\right]-\left[
\begin{array}{ccc}
(\bC_1^H\bC_1)^{-1} & 0\\
0 & 0
\end{array}
\right]\bC^H\bC+\textbf{I}\right| \label{hlp15}\\
&=&\left|\left[
\begin{array}{ccc}
(\bC_1^H\bC_1)^{-1}\mathbf{\Lambda}_1 & 0\\
0 & 0
\end{array}
\right]-\left[
\begin{array}{ccc}
\textbf{I} & (\bC_1^H\bC_1)^{-1}\bC_1^H\bC_2\\
0 & 0
\end{array}
\right]+\textbf{I}\right| \label{hlp16}\\
&=&\left|\left[
\begin{array}{ccc}
(\bC_1^H\bC_1)^{-1}\mathbf{\Lambda}_1 & -(\bC_1^H\bC_1)^{-1}\bC_1^H\bC_2\\
0 & \bI
\end{array}
\right]\right|  \nonumber \\
&=& \left|(\bC_1^H\bC_1)^{-1}\mathbf{\Lambda}_1 \right| \label{hlp17}
\end{eqnarray}
where (\ref{hlp15}) follows from the fact that $\left|\bA\bB+\bI\right|= \left|\bB\bA+\bI\right|$, and
(\ref{hlp16}) follows since
$$\bC^H\bC=\left[\bC_1 \quad \bC_2\right]^H\left[\bC_1\quad \bC_2\right]=\left[
\begin{array}{ccc}
\bC_1^H\bC_1 & \bC_1^H\bC_2\\
\bC_2^H\bC_1 & \bC_2^H\bC_2
\end{array}
\right] \; . $$

We now return to the general optimization problem in (\ref{hlp5}) with
non-zero $\bK_w$.  As the helper begins to broadcast artificial noise, both the
mutual information between the transmitter and the intended receiver
$I(X_1;Y_1)$ and the mutual information between the transmitter and
the eavesdropper $I(X_1;Y_2)$ are in general decreased.  Both of these
functions are non-increasing in $\bK_w$ since
$$\frac{\left|\bA+\bB\right|}{\left|\bB\right|}\geq
\frac{\left|\bA+\bB+\mathbf{\bigtriangleup}\right|}{\left|\bB+\mathbf{\bigtriangleup}\right|}$$
when $\bA$, $\mathbf{\bigtriangleup}\succeq 0$ and $\bB\succ 0$
\cite{Weingarten}. A favorable choice for $\bK_w$ would be one that
reduces $I(X_1;Y_2)$ more than $I(X_1;Y_1)$.  Since the optimal
solution to (\ref{hlp5}) is intractable, we propose a suboptimal approach that
introduces an additional constraint; namely, we search among those
$\bK_w$ matrices that guarantee no decrease in the favorable term
$I(X_1;Y_1)$ while the power constraint (\ref{hlp3}) is satisfied.  It should
be noted that this approach is more general than the cooperative
jamming schemes proposed in \cite{Han,Petropulu} for the MISO case
where the jamming signal is nulled out at the destination. Clearly,
such sub-optimal solutions are restricted to the case where there
exists a null space between the helper and the intended receiver.

In the following, we obtain an expression that represents all $\bK_w\succeq0$ matrices
with the power constraint $\text{Tr}(\bK_w)=P_h$ that do not impact
the mutual information between the transmitter and the intended receiver; i.e.,
$$I(X_1;Y_1)|_{\bK_w\succeq0,\bK_x=\bK_x^*}=I(X_1;Y_1)|_{\bK_w=0,\bK_x=\bK_x^*} \; ,$$
or from (\ref{hlp14})
\begin{equation}\label{hlp18}
\log\left|\bK_x^*\bH_1^H(\bG_2\bK_w\bG_2^H+\bI)^{-1}\bH_1+\bI\right|=\log\left|\bK_x^*\bH_1^H\bH_1+\bI\right|=\log\left|(\bC_1^H\bC_1)^{-1}\mathbf{\Lambda}_1\right|.
\end{equation}
Note that, without loss of generality, we have used an equality power
constraint $\text{Tr}(\bK_w)=P_h$ since for the desired $\bK_w$ the
best performance is in general obtained when helper transmits at
maximum power.

\textbf{Theorem 1:} All $\bK_w\succeq0$ matrices for which $
\log\left|\bK_x^*\bH_1^H(\bG_2\bK_w\bG_2^H+\bI)^{-1}\bH_1+\bI\right|=\log\left|\bK_x^*\bH_1^H\bH_1+\bI\right|=\log\left|(\bC_1^H\bC_1)^{-1}\mathbf{\Lambda}_1\right|$
satisfy the following relation:
\begin{equation}\label{hlp19}
\bH_1^H(\bG_2\bK_w\bG_2^H+\bI)^{-1}\bH_1=\bS^{-1/2}\left[\bC^{-H} \left[
\begin{array}{ccc}
\mathbf{\Lambda}_1 & 0\\
0 & \bN
\end{array}
\right]\bC^{-1}-\textbf{I}\right] \bS^{-1/2}
\end{equation}
where
\begin{equation}
\label{hlp20}
\begin{array}{c}
\mathbf{\Lambda}_{22} \preceq \bN\preceq\mathbf{\Lambda}_2 \\
\mathbf{\Lambda}_{22} = \bC_2^{H}\bC_2+\bC_2^H\bC_1(\mathbf{\Lambda}_1-\bC_1^H\bC_1)^{-1}\bC_1^{H}\bC_2
\end{array}
\end{equation}
and $\mathbf{\Lambda}_1$, $\mathbf{\Lambda}_2$, $\bC$, $\bC_1$ and $\bC_2$
are defined in (\ref{hlp8})-(\ref{hlp11}).

\textbf{Proof:} In Appendix A, using similar steps as those used to
obtain (\ref{hlp17}), we show that all $\mathbf{\Sigma}\succeq0$ matrices for
which
$\log\left|\bK_x^*\mathbf{\Sigma}+\bI\right|=\log\left|(\bC_1^H\bC_1)^{-1}\mathbf{\Lambda}_1\right|$
must have the following form
\begin{equation}\label{hlp21}
\mathbf{\Sigma}= \bS^{-1/2}\left[\bC^{-H} \left[
\begin{array}{ccc}
\mathbf{\Lambda}_1 & \bM\\
\bM^H & \bN
\end{array}
\right]\bC^{-1}-\textbf{I}\right] \bS^{-1/2} \; . 
\end{equation}
In the following, we obtain matrices $\bN\succeq0$ and $\bM$ and complete the proof by considering the
following specific choice for $\mathbf{\Sigma}$:
\begin{equation}\label{hlp22}
\mathbf{\Sigma}=\bH_1^H(\bG_2\bK_w\bG_2^H+\bI)^{-1}\bH_1 \; .
\end{equation}
For the specific $\mathbf{\Sigma}$ in (\ref{hlp22}), it is evident that
\begin{equation}\label{hlp23}
0\preceq\mathbf{\Sigma}\preceq \bH_1^H\bH_1.
\end{equation}
By applying the constraint $\mathbf{\Sigma}\preceq \bH_1^H\bH_1$ on (\ref{hlp21}) and using
(\ref{hlp13}), it is enough to show that:
$$\left[\begin{array}{ccc}
 \mathbf{\Lambda}_1 & \bM\\
\bM^H & \bN
\end{array}\right]\preceq \left[\begin{array}{ccc}
 \mathbf{\Lambda}_1 & 0\\
0 &  \mathbf{\Lambda}_2
\end{array}\right]$$
or equivalently that
$$\left[\begin{array}{ccc}
0 & -\bM\\
-\bM^H & \mathbf{\Lambda}_2-\bN
\end{array}\right]\succeq0 \; .$$
By applying the Schur Complement Lemma \cite{Horn}, the above relationship is true
\emph{iff} $\mathbf{\Lambda}_2-\bN\succeq0$ and $-\bM(\mathbf{\Lambda}_2-\bN)^{-1}\bM^H\succeq0$,
which in turn is true only when
\begin{eqnarray}
\bM & = & 0  \label{hlp24}\\
\mathbf{\Lambda}_2-\bN & \succeq & 0 \label{hlp25} \; .
\end{eqnarray}
Applying the results of (\ref{hlp24}) and (\ref{hlp25}) in (\ref{hlp21}) for the specific choice of
$\mathbf{\Sigma}=\bH_1^H(\bG_2\bK_w\bG_2^H+\bI)^{-1}\bH_1$, we have:
\begin{equation}\label{hlp26}
\mathbf{\Sigma}= \bS^{-1/2}\left[\bC^{-H} \left[
\begin{array}{ccc}
\mathbf{\Lambda}_1 & 0\\
0 & \bN
\end{array}
\right]\bC^{-1}-\textbf{I}\right] \bS^{-1/2} \; .
\end{equation}
Based on (\ref{hlp23}), we also need to show that $\mathbf{\Sigma}\succeq0$. From (\ref{hlp26}), it is enough to show that
$$\left[ \begin{array}{ccc}
\mathbf{\Lambda}_1 & 0\\
0 & \bN
\end{array}
\right]-\bC^{H}\bC = \left[ \begin{array}{ccc}
\mathbf{\Lambda}_1-\bC_1^{H}\bC_1 & -\bC_1^{H}\bC_2\\
-\bC_2^{H}\bC_1 & \bN-\bC_2^{H}\bC_2
\end{array}
\right]\succeq0 \; .$$
By applying the Schur Complement Lemma, the above relationship is true
\emph{iff} $\mathbf{\Lambda}_1-\bC_1^H\bC_1\succeq0$ and
$\bN-\bC_2^{H}\bC_2-\bC_2^H\bC_1(\mathbf{\Lambda}_1-\bC_1^H\bC_1)^{-1}\bC_1^{H}\bC_2\succeq0$. Using
Eqs. (\ref{hlp8})-(\ref{hlp10}), it is evident that
$$\mathbf{\Lambda}_1-\bC_1^H\bC_1=\bC_1^H\left[\bS^{\frac{1}{2}}\bH_1^H\bH_1\bS^{\frac{1}{2}}+\textbf{I}\right]\bC_1-\bC_1^H\bC_1=\bC_1^H\bS^{\frac{1}{2}}\bH_1^H\bH_1\bS^{\frac{1}{2}}\bC_1\succeq0$$
and finally the lower bound for $\bN$ is given by
$\bN\succeq\bC_2^{H}\bC_2+\bC_2^H\bC_1(\mathbf{\Lambda}_1-\bC_1^H\bC_1)^{-1}\bC_1^{H}\bC_2\succ0 \; ,$
which completes the proof.

It should be noted that as $\bN \rightarrow \mathbf{\Lambda}_{22}$,
we have Tr$(\bK_w)\rightarrow \infty$. Moreover, Tr$(\bK_w)=0$ is
achieved by $\bN=\mathbf{\Lambda}_2$. Hence, for each scalar $P_h$,
there always exists an $\bN$ in the range
$\mathbf{\Lambda}_{22}\preceq\bN\preceq\mathbf{\Lambda}_2$ that will
lead to a $\bK_w$ that satisfies (\ref{hlp19}) with Tr$(\bK_w)=P_h$.

Thus far, we have not made any assumption on the number of antennas at
each node. But it is clear from (\ref{hlp19}) that, for example when
$\bG_2$ has more columns than rows, for a fixed $\bN$ in the
acceptable range (\ref{hlp20}) there will be an infinite number of
$\bK_w$ matrices that satisfy (\ref{hlp19}) and consequently do not
decrease $I(X_1;Y_1)$. In fact, in this example, a common policy for
the helper is to simply transmit artificial noise in the null space of
$\bG_2$.  A more interesting case occurs when no such null space
exists, i.e., when the number of antennas at the helper is less than
or equal to that of the intended receiver ($n_h \leq n_r$).  The above
result demonstrates the non-trivial fact that even when $n_h \leq
n_r$, it is possible to find a non-zero jamming signal that does not
impact $I(X_1;Y_1)$ even when the jamming signal can not be nulled by
the channel.  In the next section, we find more constructive
expressions for the $\bK_w$ matrices that satisfy (\ref{hlp19}) for
various combinations of the number of antennas at different nodes.  In
particular, we show that when $n_h \le n_r$, a closed-form expression
for $\bK_w$ can be found.

\section{Results for Different Scenarios}  \label{scenarios}

In this section, we consider all possible combinations of the number
of antennas at the transmitter, helper and intended receiver, and
obtain constructive methods for computing specific $\bK_w$ matrices
that satisfy (\ref{hlp19}). Such $\bK_w$ will have no impact on
$I(X_1;Y_1)$, but will in general decrease $I(X_1;Y_2)$, the mutual
information between the transmitter and the eavesdropper, compared
with the case that there is no helper.  Hence, the secrecy level of
the confidential message is increased. As mentioned before, additional
improvement in the secrecy rate can be achieved if the transmitter
updates its covariance matrix $\bK_x$ once $\bK_w$ is computed.  Note,
however, that such an iterative process will not be pursued beyond
updating $\bK_x$; unlike the first step, where $\bK_w$ was updated
from its initial value of zero, there is no guarantee that finding a
new $\bK_w$ will reduce $I(X_1;Y_2)$. Hence, the final secrecy rate
for the proposed method is obtained by simply computing (\ref{hlp6})
and (\ref{hlp7}) for the resulting $\bK_w$ matrices derived in this
section.

\subsection{Case 1: $n_h \leq \min\{n_r,n_t\}$}\label{sec:num1}

We show here that for the case where $n_h \leq \min\{n_r,n_t\}$
and for a fixed $\bN$ in the acceptable range
(\ref{hlp20}), there is only one $\bK_w$ matrix that satisfies (\ref{hlp19}) and
consequently does not decrease $I(X_1;Y_1)$.  Using the matrix
inversion lemma, Eq. (\ref{hlp19}) can be written as:
\begin{eqnarray}
\bH_1^H(\bG_2\bK_w\bG_2^H+\bI)^{-1}\bH_1&=&\bH_1^H\bH_1-\bH_1^H\bG_2(\bG_2^H\bG_2+\bK_w^{-1})^{-1}\bG_2^H\bH_1\nonumber \\
&=&\bS^{-1/2}\left[\bC^{-H} \left[
\begin{array}{ccc}
\mathbf{\Lambda}_1 & 0\\
0 & \bN
\end{array}
\right]\bC^{-1}-\textbf{I}\right] \bS^{-1/2} \; . \nonumber
\end{eqnarray}
Replacing $\bH_1^H\bH_1$ with (\ref{hlp13}), we have:
\begin{equation}\label{hlp27}
\bH_1^H\bG_2(\bG_2^H\bG_2+\bK_w^{-1})^{-1}\bG_2^H\bH_1=
\bS^{-1/2}\bC^{-H} \left[
\begin{array}{ccc}
0 & 0\\
0 & \mathbf{\Lambda}_2-\bN
\end{array}
\right]\bC^{-1}\bS^{-1/2} \; .
\end{equation}
Since we have assumed that the channels are full rank, in the case of $n_h \leq n_r \leq n_t$ or $n_h \leq n_t \leq n_r$, it is clear that
rank$(\bG_2^H\bH_1)=n_h.$ Thus, from (\ref{hlp27}) we have:
\begin{equation}\label{hlp28}
(\bG_2^H\bG_2+\bK_w^{-1})^{-1}=
\bO^H\bS^{-1/2}\bC^{-H} \left[
\begin{array}{ccc}
0 & 0\\
0 & \mathbf{\Lambda}_2-\bN
\end{array}
\right]\bC^{-1}\bS^{-1/2}\bO
\end{equation}
where $\bO$ is the right inverse of $\bG_2^H\bH_1$, which, for example
when $n_h \leq n_r \leq n_t$, can be written as
$\bO=\bH_1^H(\bH_1\bH_1^H)^{-1}\bG_2(\bG_2^H\bG_2)^{-1}$.  The
following lemma is a direct result of Eqs.~(\ref{hlp27}) and
(\ref{hlp28}).

\textbf{Lemma 3:} For the case of $n_h \leq \min\{n_r,n_t\}$
and for a fixed $\bN$ in the acceptable range (\ref{hlp20}), the
$\bK_w\succeq0$ matrix for which (\ref{hlp19}) is satisfied and $I(X_1;Y_1)$
is not decreased is given by
\begin{equation}\label{hlp29}
\bK_w=\bQ-\bQ\bG_2^H(\bG_2\bQ\bG_2^H-\bI)^{-1}\bG_2\bQ
\end{equation}
where $\bQ$ is the RHS of (\ref{hlp28}).

\textbf{Proof:} After applying the matrix inversion lemma on the LHS of
(\ref{hlp28}), a straightforward computation yields (\ref{hlp29}).

As is evident from Eqs. (\ref{hlp28})-(\ref{hlp29}), we still have a design
parameter, $\bN$, that should be chosen in its acceptable range
$\mathbf{\Lambda}_{22} \preceq \bN \preceq \mathbf{\Lambda}_2$ such
that the power constraint $\text{Tr}(\bK_w)=P_h$ is satisfied.
Finding the optimal $\bN$ that minimizes $I(X_1;Y_2)$ when $\bK_x$
and $\bK_w$ are given by (\ref{hlp12}) and (\ref{hlp29}), respectively, is as
intractable as the general optimization problem in (\ref{hlp5}). Instead, we
simply restrict the $\bN$ we consider to those that can be linearly
parameterized within the acceptable range, as follows:
\begin{equation}\label{hlp30}
\bN=\mathbf{\Lambda}_{22}+t\left(\mathbf{\Lambda}_2-\mathbf{\Lambda}_{22}\right) \; .
\end{equation}
Consequently the term $\mathbf{\Lambda}_2-\bN$ in Eq. (\ref{hlp29}) becomes
$$\mathbf{\Lambda}_2-\bN=(1-t)\left(\mathbf{\Lambda}_2-\mathbf{\Lambda}_{22}\right)$$
where the scalar $0\leq t\leq 1$ is chosen such that the power
constraint $\text{Tr}(\bK_w)=P_h$ is satisfied. Note that as
$t\rightarrow 0$ $(\bN \rightarrow\mathbf{\Lambda}_{22})$ then
$\text{Tr}(\bK_w)\rightarrow\infty$, and as $t \rightarrow 1$
$(\bN\rightarrow\mathbf{\Lambda}_2)$ then $\text{Tr}(\bK_w)\rightarrow
0$.  Thus, we are guaranteed that an acceptable $\bN$ can be found
in this way.


\subsection{Case 2: $n_h > \min\{n_r,n_t\}$}\label{sec:num2}

As mentioned before, for the case of $n_h > n_r$ and for a fixed $\bN$
in the acceptable range (\ref{hlp20}), there are many $\bK_w$ matrices
that satisfy (\ref{hlp19}) and consequently do not decrease $I(X_1;Y_1)$.
A common policy for the helper in this case is to
transmit artificial noise in the null space of $\bG_2$. However, as
(\ref{hlp19}) shows, this policy is sufficient but it is not
necessary.  In other words, it is possible that the optimal $\bK_w$
satisfying (\ref{hlp19}) has elements outside the null space of
$\bG_2$. Because of the non-linear constraint in (\ref{hlp19}),
finding the optimal $\bK_w$ is intractable.  A similar discussion applies for the
case of $n_t < n_h \leq n_r$.

In this section, we present an approach for computing a suitable $\bK_w$.
Consider the following jamming signal covariance matrix:
\begin{equation}\label{hlpext1}
\bK_w=\mathbf{\Gamma}\, \mathbf{\Pi} \,\mathbf{\Gamma}^H \; ,
\end{equation}
where $\mathbf{\Pi}$ is a $d\times d$ positive semidefinite
matrix, and $\mathbf{\Gamma}$ is an $n_h\times d$ matrix. For the case
of $n_t < n_h \leq n_r$ or $n_h>n_r$, we can choose $\mathbf{\Gamma}$ such that
$\bG_2\,\mathbf{\Gamma}$ is orthogonal to
$\bH_1\,{\bK_x^*}^\frac{1}{2}$, i.e., ${\bK_x^*}^\frac{1}{2}
\bH_1^H\bG_2\, \mathbf{\Gamma}=\b 0$.  For example, $\mathbf{\Gamma}$
can be chosen as the $d$ right singular vectors in the nullspace of
${\bK_x^*}^\frac{1}{2} \bH_1^H\bG_2$.  Since $\bK_x$ will often be
rank deficient, the value of $d$ will typically be larger than $n_h-n_t$
for the case of $n_t < n_h \leq n_r$, and larger than $n_h-n_r$ for the case
of $n_h >n_r$.
For this choice of $\mathbf{\Gamma}$, the resulting $\bK_w$ in
(\ref{hlpext1}) satisfies (\ref{hlp19}), and doesn't decrease
$I(X_1;Y_1)$ for $\bN=\mathbf{\Lambda}_2$, as is clear from
(\ref{hlp19}).  Given $\mathbf{\Gamma}$, the choice of $\mathbf{\Pi}$
can be made to maximize the transfer of the ``information'' in the
helper's jamming signal to the eavesdropper.  In particular, note
that at the eavesdropper, the covariance of the helper's jamming
signal will be given by $\bH_2\mathbf{\Gamma\Pi\Gamma}^H\bH_2^H$. If
the eigenvalue decomposition of $\mathbf{\Gamma}^H\bH_2^H\bH_2\mathbf{\Gamma}$
is written as
$$\mathbf{\Gamma}^H\bH_2^H\bH_2\mathbf{\Gamma}=\bU\,\bD\,\bU^H$$
with $\bU$ unitary and $\bD$ square and diagonal, then $\mathbf{\Pi}$
can be found via waterfilling; i.e.,
$$\mathbf{\Pi}=\bU\,\mathbf{\Delta}\,\bU^H \; ,$$ where $\mathbf{\Delta}=\left[\eta\bI-\bD^{-1}\right]^+$,
the operation $[\bA]^+$ zeros out any negative elements, and the
water-filling level $\eta$ is chosen such that
$\text{Tr}(\bK_w)=\text{Tr}(\mathbf{\Delta})=P_h$.

\section{Numerical Results}  \label{numrst}
In this section, we present numerical results to illustrate our
theoretical findings. In all of the following figures, channels are
assumed to be quasi-static flat Rayleigh fading and independent of
each other. The channel matrices $\bH_1\in\mathbb{C}^{n_r\times n_t}$
and $\bG_2\in\mathbb{C}^{n_r\times n_h}$ have i.i.d. entries
distributed as $\mathcal{CN}(0, \sigma_{d}^2)$, while
$\bG_1\in\mathbb{C}^{n_e\times n_t}$ and
$\bH_2\in\mathbb{C}^{n_e\times n_h}$ have i.i.d. entries distributed
as $\mathcal{CN}(0, \sigma_{c}^2)$. In each figure, values for the
number of antennas at each node, as well as $\sigma_d^2$ and
$\sigma_c^2$, will be depicted. Unless otherwise indicated, results
are calculated based on an average of at least 500 independent channel
realizations.

In the first example, Fig. 1, we randomly generate positive definite
matrices $\bS$ such that $\text{Tr}(\bS)\leq P_t$. For each $\bS$, we
compute the secrecy capacity of the MIMO Gaussian wiretap channel
without helper ($\bK_w=\b0$) as given by (\ref{cs}). Next, using
(\ref{hlp29}), we obtain a $\bK_w$ with the average power constraint
$\text{Tr}(\bK_w)=P_h$ that does not decrease $I(X_1;Y_1)$, and then
update $\bK_x$ and compute $C_{sec}(\bS)$, using (\ref{hlp6}) and (\ref{hlp7}), 
accordingly.  Fig. 1 compares the secrecy capacity of
the wiretap channel with (solid lines) and without (dotted lines) the
helper.  Note that the vertical difference between the solid curves
(about 0.6 bps/channel use) represents the role of the transmit power
$P_t$ on the secrecy capacity with helper when $P_t$ changes from 100
to 150 and $P_h=20$. This relatively small difference indicates that,
in this example, $P_t$ does not have a big impact on the secrecy
capacity. Its role is even more negligible when $P_h=0$, where only an
increase of $0.3$~bps/channel use is obtained as $P_t$ increases from
100 to 150.  The role of the helper on the other hand is significantly
more important; increasing $P_h$ from 0 to 20 while holding $P_t$
fixed results in an increase on the order of 3~bps/channel use.  
Furthermore, the
use of the helper with a total power of only 120 ($P_t=100, P_h=20$)
provides significantly better secrecy performance than not using the
helper and transmitting with total power equal to 150 ($P_t=150,
P_h=0$).

In the next examples, we calculate the secrecy capacity of the
proposed algorithms under the assumption of an {\em average} power
constraint $P_t$ at the transmitter, and under the constraint that the
helper does not reduce the mutual information between the transmitter
and receiver.  While Eqs.~(\ref{hlp6}) and~(\ref{hlp7}) provide the
performance for a specific $\bS$, one must solve \cite{RLiu}, \cite[Lemma
1]{Weingarten}
\begin{equation}
\label{maxS}
C_{sec}(P_t)=\max_{\bS\succeq0, \text{Tr}(\bS)\le P_t }
C_{sec}(\bS)
\end{equation}
to find the secrecy capacity over all $\bS$ that satisfy the average
power constraint.  In the examples that follow, we perform a numerical
search to solve~(\ref{maxS}) and compute the secrecy capacity.

Fig. 2 shows the secrecy capacity versus $P_h$ for a fixed total
average power $P_t+P_h=110$. In this figure, we consider a situation
in which $\sigma_c > \sigma_d$, or in other words where
the channel between the transmitter and the intended receiver
is weaker than the channel between the transmitter and the
eavesdropper, and the channel between the helper and the intended
receiver is weaker than the channel between the helper and the
eavesdropper.  The arrow in the figure shows the secrecy capacity
without the helper $(P_h=0)$. The figure shows that a helper with just
a single antenna can provide a dramatic improvement in secrecy rate
with very little power allocated to the jamming signal; in fact, the
optimal rate is obtained when $P_h$ is less than 2\% of the total
available transmit power.  If the number of antennas at the helper
increases, a much higher secrecy rate can be obtained, but at the
expense of allocating more power to the helper and less to the signal
for the desired user.


In Fig. 3, we consider a situation in which, unlike the above example,
we have $\sigma_d > \sigma_c$.  Thus, the intended receiver, in
comparison with the eavesdropper, receives a weaker information signal
and a stronger jamming signal than the eavesdropper. It might seem
that in this situation, the helper cannot be very useful, but the
figure shows that even in this case we can have a notable improvement
in the secrecy rate (about 4~bps/channel use) by increasing the number
of antennas at the helper, and with an appropriate power assignment
between the transmitter and the helper, without requiring extra total
transmit power for the helper node.

In Fig. 4, we consider a specific scenario where the secrecy capacity in the
absence of the helper node is zero.  While channel matrices $\bH_2$ and $\bG_2$ 
are generated randomly with i.i.d. entries distributed
as $\mathcal{CN}(0, \sigma_{c}^2)$ and $\mathcal{CN}(0, \sigma_{d}^2)$, 
respectively, we assume the following specific
choices for $\bH_1$ and $\bG_1$:
$$ \bH_1=\left[\begin{array}{ccc}
-0.25 + 0.5i &  -0.35 & -1.25- 0.9i\\
-0.4 + 0.1i  &-0.2+ 0.75i &-i
\end{array}
\right] $$
$$ \bG_1=\left[\begin{array}{ccc}
2+ 0.25i &  1.5 + 0.5i &2i\\
0.25 + 0.25i  &-0.7 + 1.5i &0.5 + 0.33i\\
-1.5   &-0.5-i &- 2.9i
\end{array}
\right] . $$
Since $\bH_1^H\bH_1\preceq \bG_1^H\bG_1$, all the generalized
eigenvalues of the pencil
$$\left(\bS^{\frac{1}{2}}\bH_1^H\bH_1\bS^{\frac{1}{2}}+\textbf{I}\right)
- \gamma\left(\bS^{\frac{1}{2}}\bG_1^H\bG_1\bS^{\frac{1}{2}}+\textbf{I}\right)$$
are zero for all $\bS\succeq 0$ and consequently, the secrecy capacity
without helper will be zero.  In this example, we also assume that not
only is the total power fixed at $P_t+P_h=110$, but also the total
number of transmit antennas is fixed at $n_t+n_h=3$. As in the other
examples, the secrecy rate of the wiretap channel is considerably
improved with the helper.  In this case, the best performance is
obtained when the helper has only a single antenna.

Finally, in Fig. 5, we consider the role of number of antennas at the
helper, $n_h$, in the secrecy rate for the specific matrix power
constraint $\bS=\frac{P_t}{n_t}\bI$.  Note that the solution of
Section~\ref{sec:num1} applies for $n_h \leq 3$, while the solution
of Section~\ref{sec:num2} holds for $n_h > 3$.  In all cases, we see
that the secrecy rate increases considerably as $n_h$ increases.

\section{Conclusions}  \label{conc}
In this paper, we have studied the Gaussian MIMO Wiretap channel in
the presence of an external jammer/helper, where the helper node
assists the transmitter by sending artificial noise independent of the
source message to confuse the eavesdropper.  The jamming signal from
the helper is not required to be decoded by the intended receiver and
is treated as noise at both the intended receiver and the
eavesdropper.  We obtained a closed-form relationship for the
structure of the helper's artificial noise covariance matrix that
guarantees no decrease in the mutual information between the
transmitter and the intended receiver.  We showed how to find
appropriate solutions within this covariance matrix framework that
provide very good secrecy rate performance, even when there is no
non-trivial null space between the helper and the intended receiver.
The proposed scheme is shown to achieve a notable improvement in
secrecy rate even for a fixed average total power and a fixed total
number of antennas at the transmitter and the helper, without
requiring extra power or antennas to be allocated to the helper node.

\appendices
\section{}
We are interested in finding a relationship that represents all matrices $\mathbf{\Sigma}\succ 0$
for which
\begin{equation}\label{hlpap1}
\log\left|\bK_x^*\mathbf{\Sigma}+\bI\right|=\log\left|(\bC_1^H\bC_1)^{-1}\mathbf{\Lambda}_1\right| \; ,
\end{equation}
where
\begin{equation}\label{hlpap2}
\bK_x^*= \bS^{\frac{1}{2}}\bC \left[
\begin{array}{ccc}
(\bC_1^H\bC_1)^{-1} & 0\\
0 & 0
\end{array}
\right]\bC^H \bS^{\frac{1}{2}} \; .
\end{equation}
Using the fact that $|\bA\bB+\bI|=|\bB\bA+\bI|$, it is clear that
$\mathbf{\Sigma}$ will have the form $\mathbf{\Sigma}=\bS^{-\frac{1}{2}}\bC^{-H}\bX\bC^{-1}
\bS^{-\frac{1}{2}}$ for some matrix $\bX=\bX^H$.  Substituting this expression for
$\mathbf{\Sigma}$ into~(\ref{hlpap1}) results
in the following equation that must be solved for $\bX$:
\begin{equation}\label{hlpap3}
\log\left|\left[
\begin{array}{ccc}
(\bC_1^H\bC_1)^{-1} & 0\\
0 & 0
\end{array}
\right]\bX+\bI\right|=\log\left|(\bC_1^H\bC_1)^{-1}\mathbf{\Lambda}_1\right| \; .
\end{equation}
Write $\bX$ as $\bX=\left[
\begin{array}{ccc}
\bX_1 & \bX_2\\
\bX_2^H & \bX_3
\end{array}
\right]$ so that we have
$$\left[ \begin{array}{ccc}
(\bC_1^H\bC_1)^{-1} & 0\\
0 & 0
\end{array}
\right]\bX+\bI=\left[
\begin{array}{ccc}
(\bC_1^H\bC_1)^{-1}\bX_1+\bI & (\bC_1^H\bC_1)^{-1}\bX_2\\
0 & \bI
\end{array}
\right] \; ,$$
and note that the determinant of the above matrix is given by $\left|(\bC_1^H\bC_1)^{-1}\bX_1+\bI\right|$.
By comparing this result with (\ref{hlpap1}), we see that $\bX_1=\mathbf{\Lambda}_1-(\bC_1^H\bC_1)$. Consequently, we have:
\begin{equation}\label{hlpap4}
\mathbf{\Sigma}=\bS^{-\frac{1}{2}}\bC^{-H}\left[ \begin{array}{ccc}
\mathbf{\Lambda}_1-(\bC_1^H\bC_1) & \bX_2\\
\bX_2^H & \bX_3
\end{array}
\right]\bC^{-1} \bS^{-\frac{1}{2}}
\end{equation}
where $\bX_2$ and $\bX_3$ are still unknown and must be found as
described in the text.  It is clear that (\ref{hlpap4}) and (\ref{hlp21})
are equivalent.

\begin{figure}[t]
\begin{center}
\includegraphics[width=3.5in,height=3.5in]{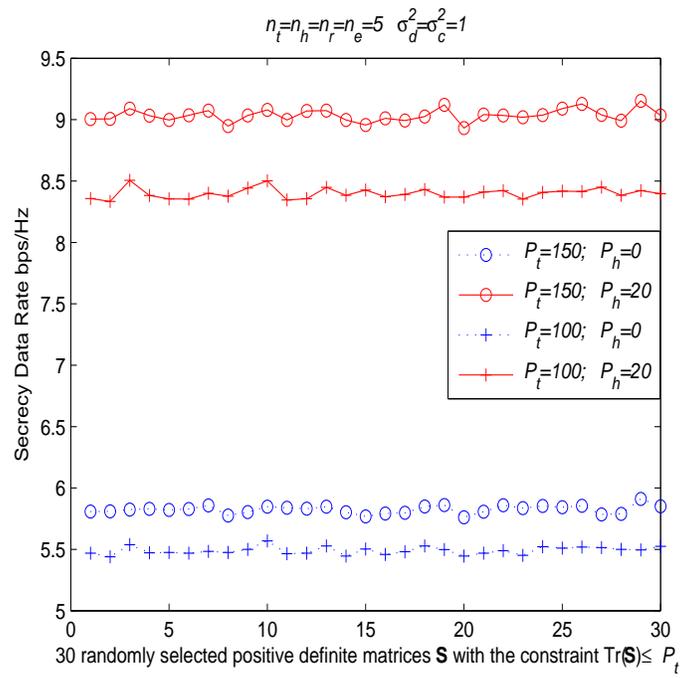}
\end{center}
\caption{Comparison of secrecy capacity for MIMO Gaussian wiretap channel with and without helper for different $P_t$ and $P_h$.}
\label{fig_sim}
\end{figure}

\begin{figure}[t]
\begin{center}
\includegraphics[width=3.5in,height=3.5in]{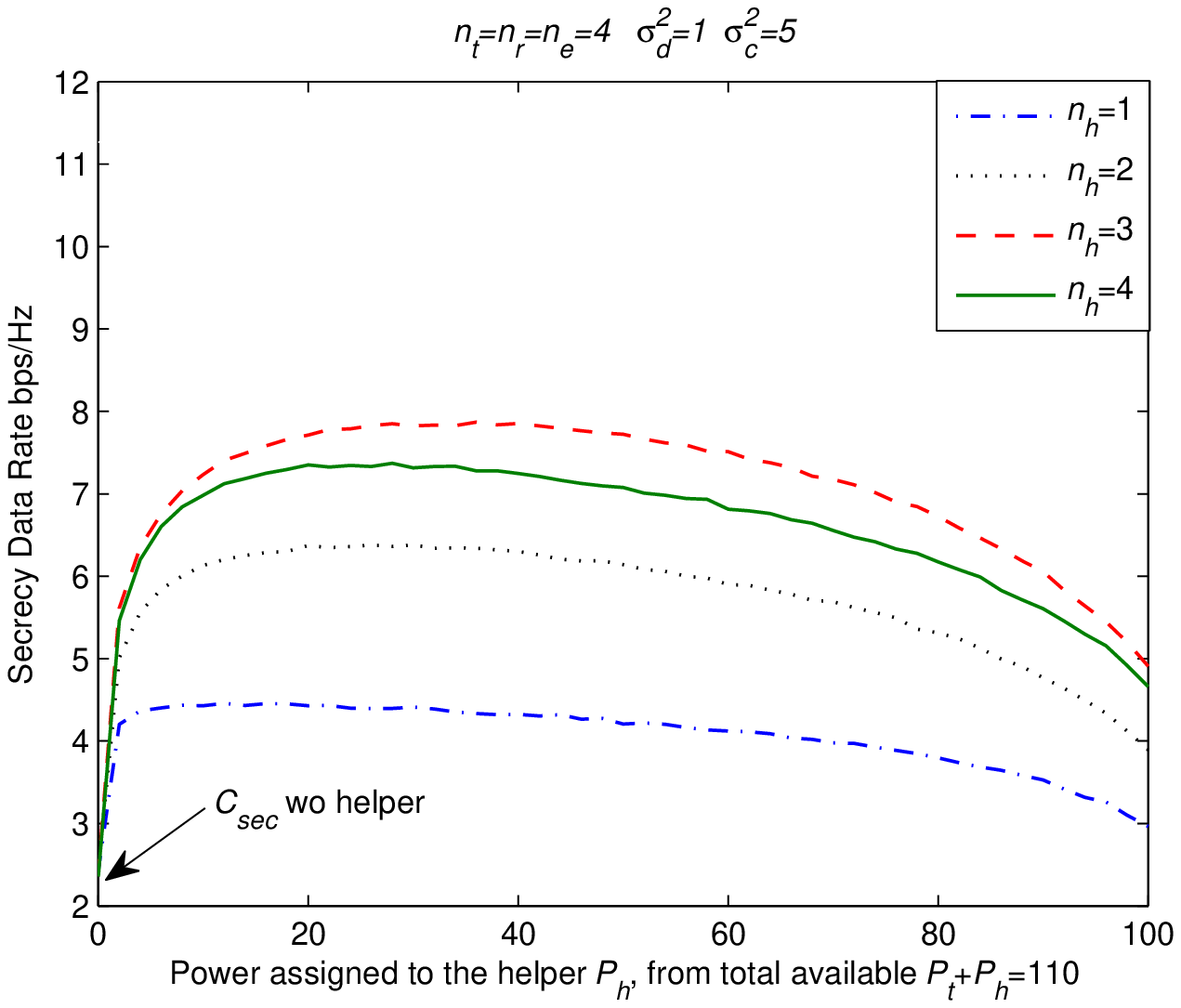}
\end{center}
\caption{Comparison of the secrecy capacity for the MIMO Gaussian wiretap channel with and without a helper versus $P_h$ for different number of
antennas at the helper, $P_t + P_h = 110$, assuming the eavesdropper's channels are stronger than those of the receiver ($\sigma_d^2=1, \sigma_c^2=5$).}
\label{fig_sim}
\end{figure}

\begin{figure}[t]
\begin{center}
\includegraphics[width=3.5in,height=3.5in]{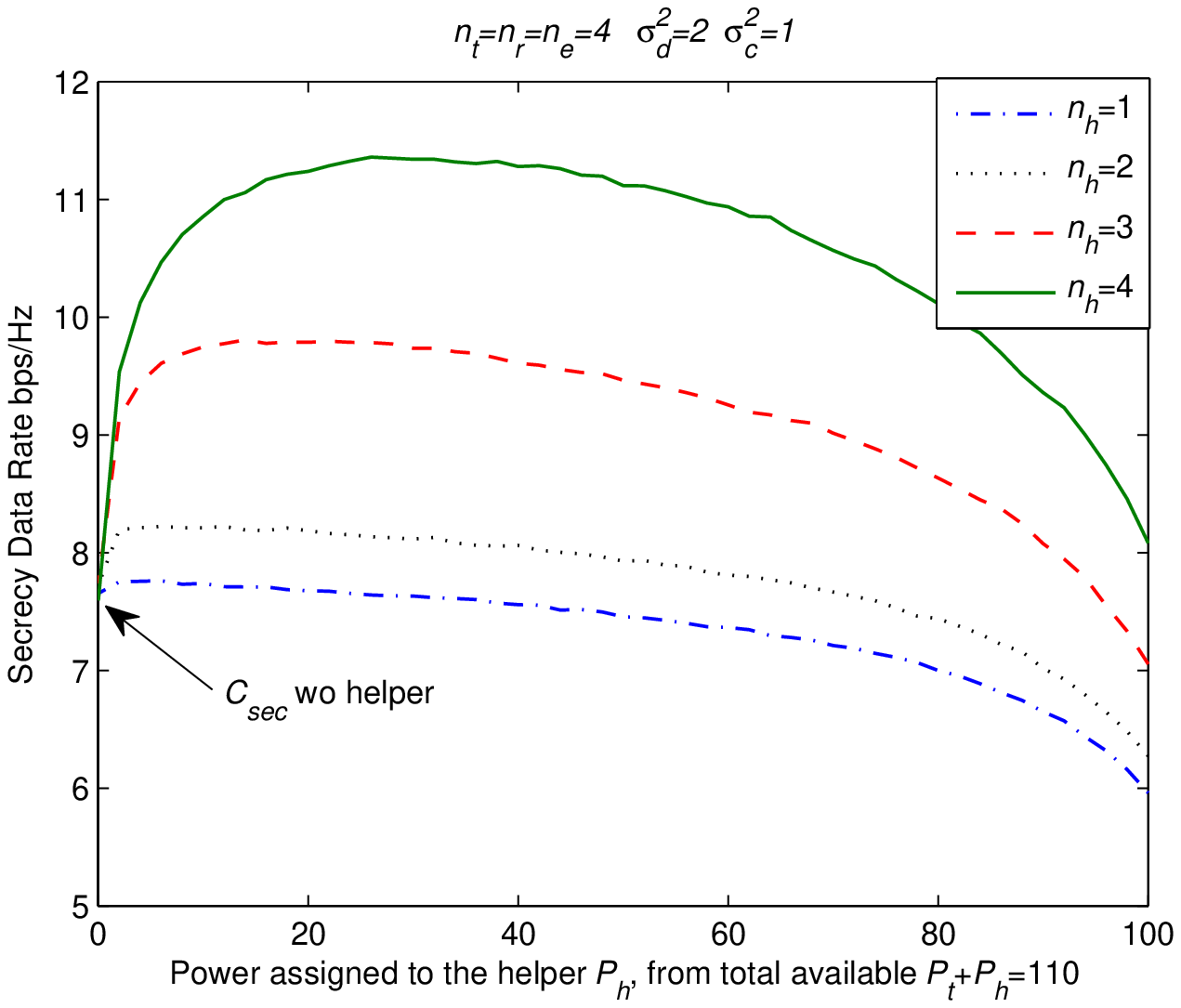}
\end{center}
\caption{Comparison of the secrecy capacity for the MIMO Gaussian wiretap channel with and without a helper versus $P_h$ for different number of
antennas at the helper, $P_t + P_h = 110$, assuming the receiver's channels are stronger than those of the eavesdropper ($\sigma_d^2=2, \sigma_c^2=1$).}
\label{fig_sim}
\end{figure}

\begin{figure}[t]
\begin{center}
\includegraphics[width=3.5in,height=3.5in]{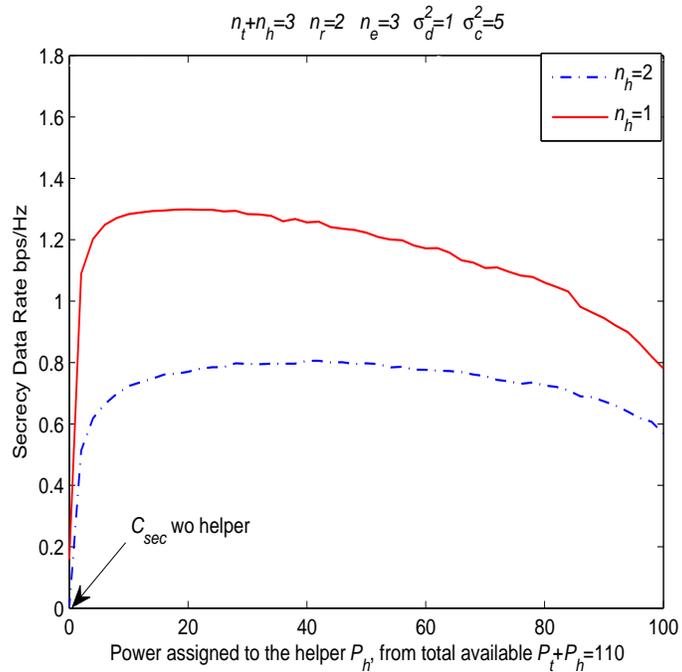}
\end{center}
\caption{Comparison of the secrecy capacity for the MIMO Gaussian wiretap channel with and without a helper versus $P_h$ for different number of
antennas at the helper, $P_t + P_h = 110$, and $n_t+n_h=3$.}
\label{fig_sim}
\end{figure}

\begin{figure}[t]
\begin{center}
\includegraphics[width=3.5in,height=3.5in]{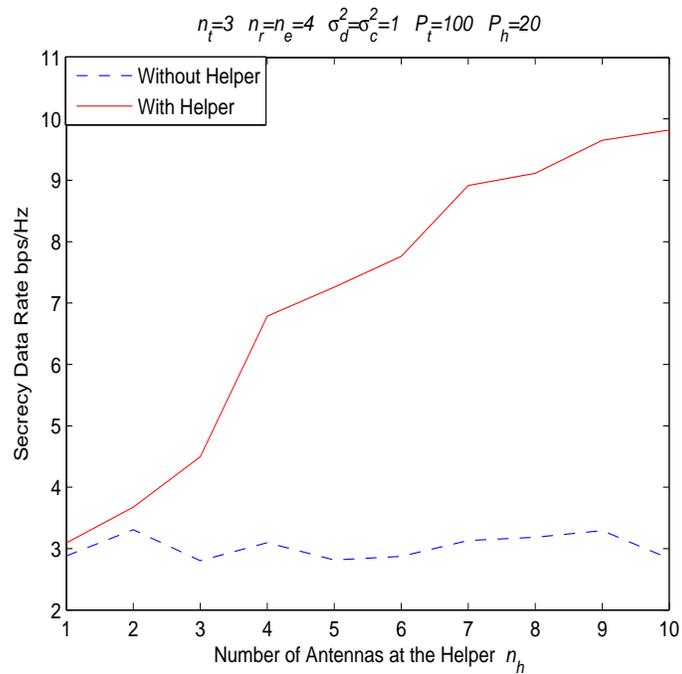}
\end{center}
\caption{Secrecy data rate versus $n_h$ for a specific matrix power constraint $\bS=\frac{P_t}{n_t}\bI$.}
\label{fig_sim}
\end{figure}
\end{document}